\documentclass[%
 reprint,
 amsmath,amssymb,
 aps,
]{revtex4-2}

\usepackage{graphicx}
\usepackage{dcolumn}
\usepackage{bm}

\usepackage{amsthm}
\usepackage{amsmath}
\usepackage{ amssymb }
\usepackage{ mathrsfs }
\usepackage{physics}
\usepackage{tikz-cd}

\newcommand{\nc}{\newcommand}
\nc{\lb}{\llbracket}
\nc{\rb}{\rrbracket}
\nc{\gl}{\llbracket}
\nc{\gr}{\rrbracket}
\nc{\bbR}{\mathbb{R}}
\nc{\bbC}{\mathbb{C}}
\nc{\bbZ}{\mathbb{Z}}
\nc{\cO}{\mathcal{O}}
\nc{\cS}{\mathcal{S}}
\nc{\cM}{\mathcal{M}}
\nc{\cT}{\mathcal{T}}
\nc{\cX}{\mathcal{X}}
\nc{\cQ}{\mathcal{Q}}
\nc{\cD}{\mathcal{D}}
\nc{\cC}{\mathcal{C}}
\nc{\cL}{\mathcal{L}}
\nc{\cG}{\mathcal{G}}
\nc{\cF}{\mathcal{F}}
\nc{\cI}{\mathcal{I}}
\nc{\cN}{\mathcal{N}}
\nc{\pd}{\partial}
\nc{\la}{\lambda}

\makeatother

\begin{document}

\preprint{APS/123-QED}

\title{Structure and Complexity of Cosmological Correlators}

\author{Thomas W.~Grimm}
\author{Arno Hoefnagels}
\author{Mick van Vliet}%

\affiliation{%
Institute for Theoretical Physics, Utrecht University,
\\
Princetonplein 5, 3584 CC Utrecht, 
The Netherlands 
}%


\begin{abstract}
\noindent
Cosmological correlators capture the spatial fluctuations imprinted during the earliest episodes of the universe. While they are generally very non-trivial functions of the kinematic variables, they are known to arise as solutions to special sets of differential equations. In this work we use this fact to uncover the underlying tame structure for such correlators and argue that they admit a well-defined notion of complexity.  
In particular, building upon the recently proposed kinematic flow algorithm, 
we show that tree-level cosmological correlators of a generic scalar field theory in an FLRW spacetime belong to the class of Pfaffian functions. 
Since Pfaffian functions admit a notion of complexity, we can give explicit bounds on the topological and computational complexity of cosmological correlators.
We conclude with some speculative comments on the general tame structures capturing all cosmological correlators and the connection between complexity and the emergence of time.
\end{abstract}

\maketitle

\section{\label{sec:intro}Introduction}

The study of the cosmological evolution of our universe gives a profound test of our fundamental theories of nature. While classical Einstein gravity coupled to matter provides a well-tested description of the late-time expansion, it is insufficient to explain the cosmological fluctuations that set the initial conditions of this evolution and explain the patterns observed in the cosmic microwave background \cite{guth_inflationary_1981,guth_fluctuations_1982,starobinsky_dynamics_1982,bardeen_spontaneous_1983,spergel_cmb_1997,hu_distinguishing_1997,dodelson_coherent_2003,mcfadden_holography_2010,mata_cmb_2013,arkani-hamed_cosmological_2015,achucarro_inflation_2022}. A proposed strategy to examine the cosmological quantum correlations is to assign a wavefunction to the universe itself which encodes the unitary time evolution of a cosmological space-time \cite{benincasa_flatspace_2018,benincasa_cosmological_2019,jazayeri_locality_2021,goodhew_cosmological_2021,dipietro_analyticity_2022,de_cosmology_2024}. The relevant cosmological correlators are obtained in the expansion of this wavefunction and evaluated on a single time-slice. 

The evaluation of the wavefunction of the universe and the extraction of the cosmological correlation functions is a computationally challenging process. In the perturbative approach \cite{maldacena_nongaussian_2003,anninos_latetime_2015,sleight_mellin_2020,arkani-hamed_cosmological_2020,benincasa_amplitudes_2022,wang_bootstrapping_2023,Arkani-Hamed:2023bsv,Arkani-Hamed:2023kig}, a large set of Feynman integrals needs to be evaluated, generally leading to an involved set of functions, which eventually reduce to a much simpler final expression for the physical amplitude. The intermediate increase in complexity becomes amplified due to the fact that the evaluation of a cosmological amplitude must involve an integral over time from past infinity to the present and take into account all possible past quantum processes. In this way, tree-level processes can already lead to non-trivial functions. It is tempting to believe that there is a less involved, indirect representation of the correlators. 

An interesting proposal for such a representation was recently put forward in \cite{Arkani-Hamed:2023bsv,Arkani-Hamed:2023kig}, were it was shown that the tree-level amplitudes of a general scalar theory with polynomial interactions in an FLRW background satisfy sets of differential equations. These equations can be found by following a rather simple set of universal rules suggesting that this approach leads to a less complex representation. However, despite the elegant kinematic flow laid out in \cite{Arkani-Hamed:2023bsv,Arkani-Hamed:2023kig} it was left open if there is an underlying structure that guarantees this reduction in complexity. 

In this work we suggest a novel way of characterizing the structure behind cosmological correlators and give a consistent mathematical framework to specify their complexity. To achieve this we will connect with some remarkable mathematical developments that originated in mathematical logic \cite{VdDries} and recently have given a foundational way to quantify the information content in a function \cite{BinNovICM,binyamini2022sharply}. While we will propose that this mathematical framework underlies general cosmological correlators, the majority of this letter is dedicated to show that it can be made concrete for the tree-level processes considered in \cite{Arkani-Hamed:2023bsv} and that it is possible to perform explicit computations. 

To summarize, we will show in this letter that the tree-level contributions to the cosmological wavefunction in a general scalar theory with polynomial interactions in an FLRW spacetime belong to a class of functions called \textit{Pfaffian functions}. This is a well-studied class of functions with several desirable properties, most notably
\begin{enumerate}
    \item[(i)] they are analytic, but satisfy numerous finiteness properties akin to polynomials;
    \item[(ii)] they are tame, made precise by the fact that they form an o-minimal structure $\bbR_{\rm Pfaff}$; 
    \item[(iii)] they admit a well-defined notion of complexity.
\end{enumerate}
Concretely, this means that  the differential equations constructed in \cite{Arkani-Hamed:2023bsv,Arkani-Hamed:2023kig} are special: they have a triangular form and are specified by polynomials. 
The theory of Pfaffian functions is thoroughly developed and we will use the data of the differential equations to give upper bounds for the \textit{topological complexity} and the \textit{computational complexity} of these functions. The former notion captures information such as the number of poles and zeros of these functions while the latter is the running time of an algorithm to check formulas satisfied by the correlators. 

Our findings allow us to make some speculations on the notion of cosmological time. In settings such as the one discussed in this paper, time can be viewed as an auxiliary concept \cite{arkani-hamed_cosmological_2020,sleight_bootstrapping_2020,baumann_cosmological_2020,pajer_building_2021,baumann_snowmass_2022,wang_bootstrapping_2023} and in \cite{Arkani-Hamed:2023bsv} the search for a more fundamental concept was initiated. Our findings now suggest a possible answer: replacing cosmological time with the computational complexity of the wavefunction of the universe. This computational complexity grows along the kinematic flow of \cite{Arkani-Hamed:2023bsv} in a quantifiable way, and we provide bounds on the complexity for generic tree-level diagrams.

Let us close by noting that it is already known that flat-space Feynman integrals are tame functions \cite{Douglas:2022ynw} and it was argued in \cite{Grimm:2023xqy} that they should also admit a complexity following a conjecture of \cite{binyamini_tameness_2023}. It is a non-trivial fact that this is also true for tree-level cosmological correlators and, even more remarkably, that this can be made fully explicit.

\section{\label{sec:CC}Cosmological Correlators and Differential Equations}
\subsection{Review of Cosmological Correlators}

In cosmology, the natural observables are in-in correlation functions given a particular initial state, arising from the fluctuations seen by an observer on a fixed time-slice. A useful formalism for analyzing these correlators uses the \textit{wavefunction of the universe} \cite{benincasa_flatspace_2018,benincasa_cosmological_2019,jazayeri_locality_2021,goodhew_cosmological_2021,dipietro_analyticity_2022,de_cosmology_2024}. Given a state of the universe $\ket{\Psi}$, this wavefunction is defined as the overlap $\Psi[\phi]=\bra{\phi}\ket{\Psi}$ with a spatial field configuration $\ket{\phi}$, and it encodes knowledge of all correlations in the state $\ket{\Psi}$. One then expands the logarithm of the wavefunction in powers of the Fourier-transformed field fluctuations, schematically
\begin{equation}
    \Psi[\phi] = \exp\Bigg(\sum_{n\geq 2} \int \dd^3 \underline{k} \,\psi_n(\underline{k}) \,\phi_{k_1}\cdots\phi_{k_n}  \Bigg) \,,
\end{equation}
where $\dd^3\underline k = \dd^3 \vec k_1 \cdots \dd^3\vec k_n$ and $\underline{k}=\{\vec{k}_1,\ldots,\vec k_n\}$. The functions $\psi_n(\underline{k})$ are the \textit{wavefunction coefficients}, and they may be computed perturbatively by means of a set of modified Feynman rules \footnote{By a slight abuse of language, we will also use the term wavefunction coefficient to refer to the contribution from a single Feynman graph.}. Given a Feynman graph contributing to a wavefunction coefficient, a convenient set of variables consists of the kinematic variables. To every vertex in the diagram one assigns a vertex energy
\begin{equation}
    X = \sum_{i} |\vec k_i|,
\end{equation}
where the sum runs over all external propagators attached to the vertex, and to every internal propagator one assigns an internal energy variable $Y$ in terms of the momenta $\vec k_i$. As an example, consider the graph

\begin{center}
\includegraphics{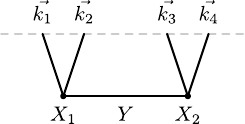} 
\end{center}

Here the dotted gray line indicates the time slice on which the state $\ket{\Psi}$ lives, and $\vec{k}_i$ are the external momenta. The kinematic variables are then given by $X_1=\vert \vec{k}_1\vert+\vert\vec{k}_2\vert$, $X_2=\vert \vec{k}_3\vert+\vert\vec{k}_4\vert$ and $Y=\vert \vec{k}_1+\vec{k}_2\vert=\vert \vec{k}_3+\vec{k}_4\vert$. In general, we collectively denote these kinematic variables $X_i$ and $Y_j$ by $Z_I$.

In this letter, we will focus on the wavefunction coefficients for a particular cosmological model. This model consists of a scalar with arbitrary polynomial interactions which is conformally coupled to gravity. Furthermore, we assume an FLRW space-time with a power-law scale factor. This theory is described by the action
\begin{equation}
    S = \int \dd^4 x \,\sqrt{-g} \left(-\frac{1}{2}(\pd \phi)^2 - \frac{1}{12}R\phi^2 -\sum_{p=3}^D \frac{\lambda_p}{p!}\phi^p\right),
\end{equation}
where the $\lambda_p$ are the couplings and $D$ is the power of the highest-order interaction in the theory. Upon infinitesimally varying the kinematic variables, one finds a set of differential equations satisfied by the wavefunction coefficients for this theory \cite{Arkani-Hamed:2023kig,Arkani-Hamed:2023bsv}.

\subsection{Kinematic Flow Algorithm}\label{ssec:kinflow}
The differential equation fulfilled by the wavefunction coefficient $\psi$ arising from a Feynman graph takes the form
\begin{equation}
    \dd I = A \, I,
\end{equation}
where $I=(\psi,f_{2},\ldots,f_{N_{\rm F}})$ is a basis vector of functions and $A$ is an $N_{\rm F}\times N_{\rm F}$ matrix of one-forms. Furthermore, an algorithm was devised which allows one to find the matrix $A$ by a combinatorial evolution procedure termed \textit{kinematic flow}. In order to understand the structure of cosmological correlators, we now provide a minimalistic review of the algorithm. For a more detailed overview we refer the reader to \cite{Arkani-Hamed:2023bsv}, and for an in-depth description and derivation we refer to \cite{Arkani-Hamed:2023kig}.

To prepare the algorithm, one first removes the external propagators, and marks the remaining edges of the graph with crosses. On the resulting marked graphs, one considers \textit{graph tubings}, which are
clusters of adjacent vertices and crosses including at least one vertex. Tubings with a single component provide a convenient representation for the \textit{letters} of the differential equation, which encode the singularity structure. The letter associated to a tubing is given by the sum of the vertex energies $X_i$ in the tube and the internal energies $Y_j$ of the edges that enter the tube. If the tube includes the cross on such an edge, the sign of the internal energy is flipped. For example, for the two-vertex the letters may be represented by the following tubings

\begin{center}
\includegraphics{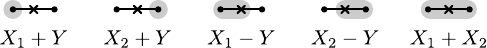} 
\end{center}

Below we now explain the steps of the algorithm, and afterwards we give an example for the two-vertex chain.
\begin{enumerate}
    \item[(i)] Select a complete tubing of the graph, for which the differential $\dd f$ of the corresponding function $f$ is to be computed.
    \item[(ii)] Write down a `kinematic flow tree' according to the following steps:
    \item[1.] Activation. For each component of the tubing, write down a descendant in which that component is `activated' (indicated by a coloring).
    \item[2.] Growth. For each activated component which contains no crosses, descendants are generated by `growing' the component by including adjacent crosses in all possible combinations.
    \item[3.] Merger. If the tube resulting from the previous step intersects another tube component, the two components merge and the union becomes activated. 
    \item[4.] Absorption. If a cross contained in an activated component is adjacent to another component containing a cross, the other component is `absorbed' whereupon the union becomes activated, generating another descendant.
    \item[(iii)] The expression for $\dd f$ is read off from the kinematic flow tree as follows: for each graph, include a term $ \dd \log \Phi$ where $\Phi$ is the letter of the active tube, and multiply this term by the  function associated to the graph minus the functions corresponding to the direct descendant graphs. Finally, multiply all terms by an overall factor $\epsilon$ and the number of vertices included in the tube upon activation.
\end{enumerate}
Applying these steps to all functions $\psi,f_2,\ldots,f_{N_{\rm F}}$ in the basis yields the matrix of one-forms $A$.

The steps of the algorithm are rather abstract, so let us apply it to a simple example, the two-vertex chain. There are four complete tubings, and hence four functions

\begin{center}
\includegraphics{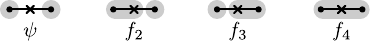} 
\end{center}

We start by computing $\dd \psi$, and write down the required kinematic flow tree below.
\begin{center}
\includegraphics{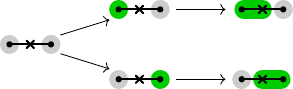} 
\end{center}

There are two components, and hence the activation step yields two descendant graphs. The next layer is generated by the growth of the active tubes to include the adjacent crosses. The kinematic flow terminates here, since the conditions for further growth, merger, or absorption are not fulfilled. Using the letters for the two-vertex chain given above, step (iii) now tells us that 
\begin{align}
    \dd \psi = \epsilon \big[ &(\psi-f_2) \dd \log(X_1+Y) + f_2 \dd\log(X_1-Y)\\ + &(\psi-f_3) \dd \log(X_2+Y) + f_3 \dd\log(X_2-Y)      \big] .\nonumber
\end{align}
Next, the kinematic flow tree for $f_2$ is given by

\begin{center}
\includegraphics{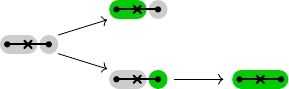} 
\end{center}

In the second layer the two components become active. The top channel terminates already at this step. The bottom channel shows the merger step, in which the entire graph is activated. The algorithm then tells us that 
\begin{align}
    \dd f_2 = \epsilon \big[ f_2 &\dd \log(X_1-Y) + (f_2-f_4)\dd \log(X_2+Y) \\ + f_4 &\dd \log(X_1+X_2)   \big] \nonumber
\end{align}
The equation for $f_3$ follows by symmetry. Finally, the kinematic flow for the final function is simply 
\begin{center}
\includegraphics{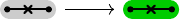} 
\end{center}

which gives (noting the factor of $2$ coming from the two vertices included in the tube upon activation)
\begin{equation}
    \dd f_4 = 2\epsilon f_4 \dd \log(X_1+X_2).
\end{equation}

\section{Cosmological Correlators as Pfaffian Functions}
\label{sec:Pfaff}

\subsection{Review of Pfaffian Functions}
Intuitively, a Pfaffian function is built from a finite sequence of functions satisfying an iterative system of differential equations, in which the derivative of any function in the chain depends polynomially on the preceding functions in the chain. More formally, let $\zeta_1,\ldots,\zeta_r:U\to\bbR$ be a finite set of functions on an open domain $U\subseteq \bbR^n$. The sequence $\zeta=(\zeta_1,\ldots,\zeta_r)$ forms a \textit{Pfaffian chain} if for each $i=1,\ldots,r$ it satisfies
\begin{equation} \label{Pfaffian_chain-ODEs}
    \dd \zeta_i = \sum_{j=1}^n P_{ij}(x_1,\ldots,x_n,\zeta_1,\ldots,\zeta_i) \,\dd x_j 
\end{equation}
where each $P_{ij}$ is a polynomial of $n+i$ variables. Note that, crucially, the polynomial $P_{ij}$ only depends on $\zeta_1,\ldots,\zeta_i$ and not on $\zeta_{i+1},\ldots,\zeta_r$. For a given Pfaffian chain, a \textit{Pfaffian function} is a function which depends polynomially on the functions in the chain, i.e.~a function of the form 
\begin{equation}
    f(x_1,\ldots,x_n) = P(x_1,\ldots,x_n,\zeta_1,\ldots,\zeta_r)
\end{equation}
where $P$ is a polynomial in $n+r$ variables.

As an example, consider the function $\zeta(x_1,\ldots,x_n)=x_1^{m_1}\cdots x_n^{m_n}$. It satisfies the differential equation 
\begin{equation}
        \dd \zeta = \sum_{j=1} m_j \zeta_j \zeta  \dd x_j,
\end{equation}
where $\zeta_j$ are the functions $\zeta_j(x_1,\ldots,x_n)=1/x_j$ for which we have
\begin{equation}
        \dd \zeta_j = -\zeta_j^2\dd x_j.
\end{equation}
In this way, the function sequence $(\zeta_1,\ldots,\zeta_n,\zeta)$ forms a Pfaffian chain.

A remarkable and powerful feature of Pfaffian functions is that they generate an \textit{o-minimal structure}, denoted by $\bbR_{\rm Pfaff}$. While a detailed exposition on o-minimal structures is beyond the scope of this letter, let us note that o-minimal structures are rooted in mathematical logic and provide a precise framework of \textit{tame geometry} \cite{VdDries}. Recently, several connections between these structures and quantum field theory were explored \cite{Grimm:2021vpn,Douglas:2022ynw,Douglas:2023fcg,Grimm:2023xqy}.

\subsection{Pfaffian Chains for Cosmological Correlators}

We will now argue that the basis functions discussed in the previous section form a Pfaffian chain. As a preparation, we need to include the $\dd\log$ terms of the letters into the chain. For each letter $\Phi_i(Z_I)= \sum_{I} c_i^I Z_I $, we have
\begin{equation}
    \dd\log \Phi_i = \frac{1}{\Phi_i}\sum_I c_i^I\dd Z_I.
\end{equation}
The coefficient $\ell_i=1/\Phi_i$ can be realized as a Pfaffian function by means of the differential equation
\begin{equation}
    \dd \ell_i = -\sum_{I}\ell_i^2 c_i^I \dd Z_I .
\end{equation}
Thus, the first part of the Pfaffian chain for cosmological correlators consists of the functions $(\ell_1,\ldots,\ell_{N_{\rm L}})$.

To set up a chain for the basis functions in the kinematic flow, we first organize them by the number of \textit{cuts} (line segments between vertices and crosses) appearing in the complete graph tubing. Complete tubings of a graph may then be uniquely described by the placement of these cuts. 

The crucial observation is that, as a consequence of the rules of the algorithm, the differential of every basis function only depends on the function itself and basis functions corresponding to complete tubings with \textit{strictly fewer} cuts. In particular, the growing of the tubes in step 2.-4.~always implies a reduction in the number of cuts. Meanwhile, there is a single basis function $\zeta_0$ which corresponds to zero cuts, i.e.~the complete tubing with one component. Taking this as the first basis function in the chain we can iteratively add the functions with an increasing number of cuts, thereby obtaining a Pfaffian chain. To be precise, let us denote the basis functions by $\zeta_{i,j}$, where $i$ is the number of cuts and $j=1,\ldots,n_i$ enumerates the basis functions with $i$ cuts. By the above argument we then have
\begin{equation}
    \dd \zeta_{i,j} = \sum_{I}  P_I (\ell_1,\ldots,\ell_{N_{\rm L}},\zeta_{i,j},\zeta_{i-1,1},\ldots)\dd Z_I \, ,
\end{equation}
which precisely fits the defining condition of a Pfaffian chain.

Let us again turn to the two-vertex chain to exemplify our argument. The inverse letters comprise the first five functions of the chain. For instance, one has $\ell_1=1/(X_1+Y)$, for which the required Pfaffian differential equation is given by $\dd\ell_1 = -\ell_1^2 \dd X_1 -\ell_1^2 \dd Y$.
Now we are able to write down the differential equations satisfied by the basis functions $(\psi,f_2,f_3,f_4)\equiv(\psi,\zeta_{1,1},\zeta_{1,2},\zeta_0)$ in terms of a Pfaffian chain, starting with $f_4$, which satisfies
\begin{align}
    \dd \zeta_0 &= 2\epsilon \ell_5 \zeta_0 (\dd X_1 +\dd Y) \,.
\end{align}
Next, we define the basis functions belonging to graphs with one cut, namely $\zeta_{1,1}$ and $\zeta_{1,2}$. Employing the algorithm, one finds that they satisfy the Pfaffian equations
\begin{align}
     \dd\zeta_{1,1} =& \,\epsilon \ell_2 \zeta_{1,1}(\dd X_1-\dd Y) +\epsilon \ell_5 \zeta_0(\dd X_1 +\dd X_2) \\
     &+ \epsilon \ell_3(\zeta_{1,1}-\zeta_0)(\dd X_2 +\dd Y)\,, \nonumber\\ 
     \dd\zeta_{1,2} =& \,\epsilon \ell_4 \zeta_{1,2}(\dd X_2-\dd Y) +\epsilon \ell_5 \zeta_0(\dd X_1 +\dd X_2)\nonumber \\
     &+ \epsilon \ell_3(\zeta_{1,2}-\zeta_0)(\dd X_1 +\dd Y)\,. \nonumber
\end{align}
Finally, the wavefunction itself is recovered in the Pfaffian chain as
\begin{align*}
\dd \psi=\,\, &\epsilon \ell_1(\psi-\zeta_{1,1})(\dd X_1 + \dd Y) + \ell_2 \zeta_{1,1} (\dd X_1-\dd Y)  \nonumber \\
 + &\epsilon \ell_3(\psi-\zeta_{1,2})(\dd X_2 + \dd Y) + \ell_4 \zeta_{1,2} (\dd X_2-\dd Y) \,. \nonumber \\
\end{align*}

\section{\label{sec:Complexity}Complexity and Emergent Time}

\subsection{Pfaffian Complexity of Cosmological Correlators}

Pfaffian functions admit a well-defined notion of complexity defined from the underlying Pfaffian chain. It is characterized by four numbers, namely the number of variables, $n$; the length of the chain, $r$ (also called the \textit{order}); the degree of the chain, $\alpha$, defined by $\alpha=\max_{i,j}(\deg(P_{ij}))$; and the degree of the Pfaffian function, $\beta$, defined by $\beta=\deg(P)$. Together, these comprise the \textit{Pfaffian complexity} of the function, denoted by
\begin{equation} \label{PfaffC}
    \cC(f) = (n,r,\alpha,\beta).
\end{equation}
The interpretation of the Pfaffian complexity will be discussed in the next subsection, but let us already note that fundamentally, it provides a measure of logical information which is required to specify a function.

The results from the previous section therefore allow us to estimate the complexity of a wavefunction coefficient in terms of the number of vertices of the graph, $N_{\rm V}$. The number of variables in the chain, i.e.~the number of kinematic variables, is given by $n=2N_{\rm V}-1$, since there is one variable for each vertex and edge. The order of the chain, $r$, is the component of the Pfaffian complexity which is most strongly dependent on the number of vertices. For a given graph, the Pfaffian chain consists of $N_{\rm L}$ letters and $N_{\rm F}$ basis functions. The number of letters strongly depends on the topology of the graph. For instance, we have
\begin{align} \label{N-bounds}
    N_{\rm L}^{\rm chain}(N_{\rm V})&= 2N_{\rm V}^2 -2N_{\rm V}+1, \\
    N_{\rm L}^{\rm star}(N_{\rm V}) &= 3^{N_{\rm V}-1} +2N_{\rm V}-3.    \nonumber
\end{align}
It is worth noting that for fixed $N_{\rm V}$, the chain graph has the least letters and the star graph has the most letters, i.e.~we always have 
\begin{equation} \label{N_Lbound}
   N_{\rm L}^{\rm chain}(N_{\rm V}) \leq N_{\rm L}\leq N_{\rm L}^{\rm star}(N_V)\ . 
\end{equation}
Note that a star graph with $k$ legs is only present when the theory under consideration contains a $\phi^k$ interaction. In particular, in the notation of Section \ref{sec:CC} we require $k\leq D$ and $\lambda_k\neq 0$. Meanwhile, counting the number of basis functions one finds that
\begin{equation}
    N_{\rm F} = 4^{N_{\rm V}-1}.
\end{equation}
Therefore, for large graphs, we have $r\sim 4^{N_{\rm V}}$. The degrees of the Pfaffian chain are independent of $N_{\rm V}$, since the system of differential equations is linear in the basis functions and the coefficients are given by the inverse letters $\ell_i$. In particular, we have $\alpha=2$ and $\beta=1$, so the complexity of an $N_{\rm V}$-vertex wavefunction coefficient is 
\begin{equation} \label{PfaffC-ex}
    \cC(\psi) = (2N_{\rm V}-1, 4^{N_{\rm V}-1} + N_{\rm L},2,1).
\end{equation}
The dependence of the complexity as a function of $N_{\rm V}$ is most apparent in the order $r$ of the chain, which shows an exponential growth. Also, note that the order $r$ increases by one for every function added to the Pfaffian chain. Since the steps of the kinematic flow algorithm generate more functions to be added to the chain, we note that \textit{complexity grows along the kinematic flow in a quantifiable way}. The following subsections are devoted to the interpretation of these observations.

Let us note that the preceding discussion refers to the contribution to the wavefunction coefficient coming from a single Feynman graph viewed as a function of the kinematic variables $Z_I$. In general, the full wavefunction coefficients $\psi_n$ arise from a sum of graphs, and it will be a function of the couplings and kinematic variables $\psi_n(Z_I,\lambda_p)$. However, at tree level the sum over graphs is finite, implying that the tame structure in $Z_I$ is preserved while the coupling dependence is now a polynomial in the $\lambda_p$ with $p\leq n$. The complexity of the full tree level wavefunction coefficient viewed as Pfaffian function in $Z_I$ and $\lambda_p$ can then be obtained through the complexities of the underlying graphs which are determined by the interaction terms in the Lagrangians. It is an exciting task to study possible reductions of this total complexity, particularly due to relations or symmetries in the Lagrangian. 

\subsection{Minimizing Complexity: Lessons from Topology}
What is the meaning of the Pfaffian complexity introduced in the previous subsection? The Pfaffian complexity $\mathcal{C}(f)$ given in \eqref{PfaffC} can be viewed as giving a measure of how much information is minimally needed to define the function $f$. It may at first seem peculiar that one needs to specify four numbers to describe this measure of information, but it turns out that a single number is is not rich enough to specify the information content of a function \cite{binyamini2022sharply}. 
Crucially, the Pfaffian complexity also depends on the precise representation of the function, and one might be able to reduce the complexity by making clever choices, for example, for the Pfaffian chain. As explained in more detail in \cite{Grimm:2023xqy} this freedom is a key feature of this consistent notion of complexity and pertains beyond the Pfaffian context. Hence, our result \eqref{PfaffC-ex} gives an upper bound on the information content. There are thus two natural questions to ask:
\begin{itemize}
\item[(1)] What characterizes a minimally complex representation of $\psi$? How should one minimize the list of integers $\cC(\psi)$?
\item[(2)] What is the physical interpretation of the minimal complexity of $\psi$?
\end{itemize}
The question (1) will be addressed below, while we leave speculations on the answer to (2) to the next subsection. 

One of the essential features of Pfaffian complexity is its ability to encode other notions of complexity \cite{GabVor04}. Specifically, as alluded to in the introduction, Pfaffian complexity encodes topological and computational complexity. Though we will use these terms rather loosely, the precise meaning is that a geometric object constructed from Pfaffian functions obeys complexity bounds which are controlled by the underlying Pfaffian complexity. As a basic example, a Pfaffian function $f$ with complexity $\cC(f)=(1,r,\alpha,\beta)$ has a bound on the number of isolated zeros given by
\begin{equation}\label{eq:Bezout}
    \# \{x\,|\,f(x)=0 \} \leq 2^{r(r-1)} \beta \big(\alpha+\beta\big)^r \,,
\end{equation}
which may be interpreted as a generalized B\'ezout bound~\cite{GabVor04}. Since it counts the number of connected components of the solution set of an equation, it provides a coarse measure of topological complexity. Let us note that this example only scratches the surface of the rich theory of Pfaffian complexity and one can determine the complexity of much more involved sets \cite{GabVor04}.
These more involved examples and applications of these ideas to physical settings are discussed in \cite{Grimm:2023xqy}.

Consequently, a representation which minimizes complexity could be one which minimizes the bound on one of the derived notions of complexity, such as the topological complexity given above. However, one characteristic of all the complexity bounds derived in \cite{binyamini2022sharply} is that they grow exponentially in $r$, the order of the chain. In our analysis of cosmological correlators, we find that $r$ itself grows exponentially in $N_{\rm V}$ the number of vertices of the Feynman graph, leading to a doubly exponential growth in the topological or computational complexity of cosmological correlators. The bound given in equation \eqref{eq:Bezout} applied to the wavefunction coefficient of an $N_{\rm V}$-vertex graph as a function of the kinematic variables becomes
\begin{equation} \label{BezoutCC}
    2^{(4^{N_{\text{V}}-1} +N_{\rm L})(4^{N_{\rm V}-1} +N_{\rm L}-1) }\big(4N_{\rm V}-1\big)^{4^{N_{\rm V}-1} +N_{\rm L}}  \,.
\end{equation}
Note that the number of letters $N_{\rm L}$ is bounded in terms of $N_{\rm V}$ as in \eqref{N_Lbound} with \eqref{N-bounds} implying  doubly exponential grown in $N_{\rm V}$. 

The expression \eqref{BezoutCC} also gives a bound on the 
poles of $\psi$, since we can equally consider $1/\psi$ in \eqref{eq:Bezout} by minimally increasing the chain. However, from the integral representation of the cosmological correlators one would expect that the number of physical poles is only single-exponentially related to the number of letters and hence grows much slower with $N_{\rm V}$. Hence, we conclude that the representation of the cosmological correlators with the Pfaffian chain found in section \ref{sec:Pfaff} using the kinematic flow algorithm is far from optimal. It is apparent that the fast growth arises from the exponential growth of length of the chain with $N_{\rm V}$. In order to minimize the topological complexity, one would ideally like to find a representation that has a slowly growing chain, with maximally exponentially growing degrees $\alpha,\beta$ of the polynomials defining it. There is also a physical argument for this reduction of complexity. While a general solution to the differential equations may have a complicated singularity structure, many of these are not singularities of the wavefunction coefficient \cite{arkani-hamed_cosmological_2017,salcedo_analytic_2023,Caloro:2023cep,lee_amplitudes_2024,fan_cosmological_2024}. Therefore, the true topological complexity should be smaller than estimated here.

A guiding principle for finding a simpler representation  arises from the locality of the physical theory. It was shown in \cite{Arkani-Hamed:2023kig} that, for some specific examples, locality implies the existence of simpler sets of differential equations for the cosmological correlators. From these one can then find a shorter Pfaffian chain than initially expected. It will be shown in an upcoming work \cite{futureworkGH} 
that there is a general underlying mathematical structure
responsible for the this simplification. We expect that this can universally lead to a reduction in complexity and aim to make this precise in future work. 

\subsection{Computational Complexity and Emergent Time} \label{sec:emergenceoftime}

Another measure of complexity derived from Pfaffian complexity is computational complexity. In general, this is a quantity which measures how complicated it is to algorithmically compute a geometric object built from Pfaffian functions. The algorithms in question are based on \textit{real numbers machines}, which are computational devices capable of performing exact computations on real numbers. If $X$ is a $d$-dimensional set defined by $M$ Pfaffian equalities or inequalities of complexity $(n,r,\alpha,\beta)$, then the computational complexity of $X$ is estimated by \cite{GabVor04} 
\begin{equation}
   M^{(r+n)^{O(d)}}(\alpha+\beta)^{(r+n)^{O(d^2n)}}. 
\end{equation}
As a special case, one may consider the complexity of computing the zeros of a Pfaffian function, which is estimated by
\begin{equation}
    (\alpha+\beta)^{(r+n)^{O(d^2n)}}
\end{equation}
In the present setting, this may be interpreted as the computational complexity of the wavefunction coefficient attaining a certain prescribed value, and using the result of equation \eqref{PfaffC-ex}, we find the bound
\begin{equation}
     3^{(4^{N_{\rm V}}+ N_{\rm L}+ 2N_{\rm V}-1)^{d^2(N_{\rm L}-1)}}\,.
\end{equation}
Here $d$ is the dimension of the locus on which $\psi$ attains a certain value $\psi_0$.

The authors of \cite{Arkani-Hamed:2023bsv,Arkani-Hamed:2023kig} argue that the kinematic flow algorithm is a boundary avatar of the cosmological time evolution in the bulk. In this static description, time arises as an emergent concept. The discussion above tells us that there is a quantifiable growth of computational complexity along the kinematic flow. It is therefore tempting to speculate on the connection between time and complexity, and the idea that complexity may provide an emergent description of time. In physics there are many hints that time and complexity are related, for instance through entropy, or more recently the idea of holographic complexity and time evolution in the bulk \cite{Susskind:2014rva,Carmi:2017jqz,Susskind:2018pmk}. The notion of complexity used here is somewhat different than the one used in these works, and thereby provides a complementary perspective. Building further upon the idea of emergent time through kinematic flow, our findings provide a possible step in giving a quantitative connection between complexity and time, through the computational complexity of wavefunction coefficients. It is a compelling idea whether the aforementioned notion of computational complexity has a physical interpretation in which the universe acts as a computational device. We emphasize that these are speculative comments, and leave a further exploration of these ideas open to future work.

\section{\label{sec:conclusion}Discussion and Outlook}

In this work we have shown that that the tree-level amplitudes for a general scalar theory with polynomial interactions in an FLRW background are Pfaffian functions in the kinematic variables. This implies that they are polynomials in these variables as well as a finite set of basis functions. 
Furthermore, these basis functions must satisfy a set of differential equations \eqref{Pfaffian_chain-ODEs} of a special `triangular' form. This formulation allowed us to record their information content, or Pfaffian complexity, which we in turn used to give bounds on the topological and computational complexities of the correlators. Working with Pfaffian functions resamples several aspects of working with polynomials while at the same time including non-trivial functions as building blocks. Our results were fully explicit and made use of the algorithm of \cite{Arkani-Hamed:2023kig}. 

The fact that there is a notion of complexity for Pfaffian functions allowed us to derive explicit numerical bounds on the topological complexity, e.g.~giving a bound on the number of poles and zeros, and their computational complexity. Our result suggests, however, that the representation of cosmological correlators using the kinematic flow algorithm of \cite{Arkani-Hamed:2023kig} is not optimal. To find a more optimal representation of the same physical correlators is a crucial task for the future and will be addressed in \cite{futureworkGH,futureworkGHV}.

We have focused on the complexity associated to an individual graph. An immediate follow-up question is to inquire about the complexity of the full tree-level wave function coefficient, 
both as a function of the kinematic variables and coupling constants. 
This complexity will depend on the properties of the Lagrangian, such as the number of scalar fields, the number of non-vanishing coupling constants, and the precise form of  the interaction polynomial.  Interestingly, the Pfaffian framework allows us to quantify this dependence for tree-level amplitudes.
Furthermore, the framework allows us to quantify how simplicity emerges in amplitudes, since it accounts for possible algebraic relations among functions and gives a complexity reduction in the presence of a symmetry. An example of this arises
in \cite{Arkani-Hamed:2023kig} when considering multiple fields for which the basis functions are shared between different channels. Due to these relations, the total complexity will be lower than expected from an analysis of the individual diagrams. Investigating these possibilities is a key application of this notion of complexity, especially when eventually considering the full expression of the cosmological wave function. 

A foundational conclusion of our findings is that cosmological correlators, at least at the perturbative level, have two core features: (1) they are functions definable in an o-minimal structure, (2) they admit a notion of complexity. For the tree-level correlators we have identified this structure explicitly, namely the structure $\bbR_{\rm Pfaff}$ generated by all Pfaffian functions. $\bbR_{\rm Pfaff}$ is the benchmark example of what is now known as sharply o-minimal structure \cite{BinNovICM,binyamini2022sharply,[{Strictly speaking the sharply o-minimality of $\bbR_{\rm Pfaff}$ is only proved when restricting the functions to smaller domains and conjectured in general \cite{binyamini_tameness_2023}}] emptycit}. This notion of sharp o-minimality \text{($\sharp$o-minimality)} means that all functions, or sets, have a notion of sharp complexity (see \cite{Grimm:2023xqy} for a comprehensive discussion).  
We believe that this general notion suffices to describe all correlators and hence propose:\\[.2cm]
\textit{All perturbative cosmological correlators, viewed as functions of the kinematic variables, are definable in a sharply o-minimal structure and therefore have a notion of complexity.}\\[-.2cm]

This general conjecture can be motivated by the fact that ordinary Feynman integrals have been shown in \cite{Douglas:2022ynw} to be definable in an o-minimal structure using their relation to period integrals. Furthermore, all period integrals are expected to be sharply o-minimal by a mathematical conjecture \cite{BinNovICM}, leading to the proposal that all Feynman amplitudes are sharply o-minimal \cite{Grimm:2023xqy}. The above statement extends this to cosmological correlators. It is instructive to note that, in general, integration does not preserve the tameness property of a function and therefore the statements above are non-trivial. Establishing the above conjecture would open the door to use many of the remarkable recent results on o-minimal structures to obtain, for example, general statements about algebraic relations among amplitudes.

\begin{acknowledgments}
We would like to thank Daniel Baumann, Gal Binyamini, Lou van den Dries, Michael Douglas, Guilherme Pimentel, Lorenz Schlechter, and Stefan Vandoren for useful discussions and comments. This research is supported, in part, by the Dutch Research Council (NWO) via a Vici grant.
\end{acknowledgments}


\bibliography{apssamp}

\end{document}